\documentclass[preprint]{revtex4}
\usepackage{here}
\usepackage{amsmath}
\usepackage{amssymb}
\usepackage{graphicx}
\usepackage{amsfonts}

\begin{document}

\title{Observability of the Higgs Boson in the Presence of
Extra Standard Model Families at the Tevatron}
\author{E. Arik}
\affiliation{Bo\u gazi\c ci University, Faculty of Arts and
Sciences, Department of Physics, 34342 Bebek, Istanbul, Turkey}
\author{O. \c Cak\i r}
\affiliation{Ankara University, Faculty of Sciences, Department of
Physics, 06100 Tandogan, Ankara,  Turkey}
\author{S. A. \c Cetin}
\affiliation{Do\u gu{\c s}  University, Faculty of Arts and
Sciences, Department of Math and Sciences, 34722 Ac\i badem -
Kad{\i}k\" oy, Istanbul, Turkey}
\author{S. Sultansoy}
\affiliation{Gazi University, Department of Physics, 06500
    Teknikokullar, Ankara, Turkey\\
Institute of Physics, Academy of Sciences, H. Cavid Ave. 33, 370143
Baku, Azerbaijan}

\begin{abstract}
The observability of the Higgs boson via the $WW^{*}$ decay channel
at the Tevatron is discussed taking into account the enhancements
due to the possible existence of the extra standard model (SM)
families. It seems that the existence of new SM families can give
the Tevatron experiments (D0 and CDF) the opportunity to observe the
intermediate mass Higgs boson before the LHC.
\end{abstract}

\pacs{13.85.Rm, 14.80.Bn, 14.80.-j}

\maketitle

\section{Introduction}

It is known that the number of fermion generations is not fixed by
the standard model (SM). Asymptotic freedom of the quantum
chromodynamics (QCD) suggests that this number is less than eight.
Concerning the leptonic sector, the large electron positron collider
(LEP) data determine the number of light neutrinos to be N$=2.994\pm
0.012$ [1]. On the other hand, the flavor democracy (i.e. democratic
mass matrix approach [2-5]) favors the existence of the fourth SM
family [6-9].

Direct searches for the new leptons ($\nu_4, \ell_4$) and quarks
($u_4, d_4$) led to the following lower bounds on their masses [1]:
$m_{\ell_4} > 100.8$ GeV; $m_{\nu_4} > 45$ GeV (Dirac type) and
$m_{\nu_4} > 39.5$ GeV (Majorana type) for stable neutrinos;
$m_{\nu_4} > 90.3$ GeV (Dirac type) and $m_{\nu_4} > 80.5$ GeV
(Majorana type) for unstable neutrinos; $m_{d_4}
> 199$ GeV (neutral current decays), $m_{d_4} > 128$ GeV (charged
current decays). The precision electroweak data does not exclude the
fourth SM family, even a fifth or sixth SM family is allowed
provided that the masses of their neutrinos are about 50 GeV [14,
15].

In the Standard Model, the Higgs boson is crucial for the
understanding of the electroweak symmetry breaking and the mass
generation for the gauge bosons and the fermions. Direct searches at
the CERN $e^+e^-$ collider (LEP) yielded a lower limit for the Higgs
boson mass of $m_H > 114.4$ GeV  at 95\% confidence level (C.L.)
[1].

In this study, we present the observability of the Higgs boson at
the Tevatron and find the accessible mass limits for the Higgs boson
in the presence of extra SM fermion families (SM-4, SM-5 and SM-6).

\section{Anticipation for the Fourth SM Family}

According to the SM with three families, before the spontaneous
symmetry breaking, quarks are grouped into the following SU(2) x
U(1) multiplets:
\begin{eqnarray}
\left( \begin{array} {c} {u_L^0} \\ {d_L^0} \end{array} \right)
u_R^0\, , \, d_R^0  \qquad
\left( \begin{array} {c} {c_L^0} \\ {s_L^0} \end{array} \right)
c_R^0\, , \, s_R^0  \qquad
\left( \begin{array} {c} {t_L^0} \\ {b_L^0} \end{array} \right)
t_R^0\, , \, b_R^0  \qquad
\end{eqnarray}
where $0$ denotes the SM basis. In one family case, e.g. $d$-quark mass
is obtained due to the Yukawa interaction
\begin{eqnarray}
L_Y^{(d)} = a^d \, \left(\bar u_L \, \bar d_L  \right)
\left( \begin{array} {c} {\phi^+} \\ {\phi^0} \end{array} \right) \,
d_{R} + h.c.
\end{eqnarray}
which yields
\begin{eqnarray}
L_m^{(d)} = m^d \bar d d
\end{eqnarray}
where $m^d = a^d \eta/\sqrt{2}$ and $\eta = 2\, m_W/g_W = 1 /\sqrt{\sqrt{2} \, G_F}
\approx 246$ GeV.  In the same manner, $m^u = a^u \eta/\sqrt{2}$,
$m^e = a^e \eta/\sqrt{2}$ and
$m^{\nu_e} = a^{\nu_e} \eta/\sqrt{2}$ if $ {\nu_e}$ is a Dirac particle.

In $n$-family case
\begin{eqnarray}
L_Y^{(d)} = \sum_{i,j = 1}^n a^d_{ij} \, \left(\bar u^0_{Li} \, \bar
d^0_{Li}  \right) \left( \begin{array} {c} {\phi^+} \\ {\phi^0}
\end{array} \right) \, d^0_{Rj} + h.c. \Rightarrow \sum_{i,j = 1}^n
m^d_{ij} \, \bar d^0_{i}d^0_{j}
\end{eqnarray}
where $d_1^0$ denotes $d^0$, $d_2^0$ denotes $s^0$ etc. and
$m_{ij}^d \equiv a_{ij}^d \eta/\sqrt{2}$.

Before the spontaneous symmetry breaking, all quarks are massless
and there are no differences between $d^0$, $s^0$, $b^0$, etc. In
other words, fermions with the same quantum numbers are
indistinguishable. This leads us to the first assumption [2, 3]:
\begin{itemize}
\item{Yukawa couplings are equal within each  type of fermion families}
\end{itemize}
\begin{eqnarray}
a^d_{ij} \approx a^d \, , \, a^u_{ij} \approx a^u \, , \,
a^{\ell}_{ij} \approx a^{\ell} \, , \, a^{\nu}_{ij} \approx a^{\nu} \, . \,
\end{eqnarray}

The first assumption results in $n-1$ massless particles and one
massive particle with $m = n a^F \eta/\sqrt{2} \, (F = u, d, \ell, \nu)$ for
each type of fermion $F$.
If there is only one Higgs doublet which gives Dirac masses to all
four types of fermions ($u, d, \ell, \nu)$,
it seems natural to make the second assumption [6, 8]:
\begin{itemize}
\item{Yukawa couplings for different types of fermions should be
  nearly equal}
\end{itemize}
\begin{eqnarray}
a^d \approx a^u  \approx a^{\ell}  \approx a^{\nu}  \approx a \, .
\end{eqnarray}
Considering the mass values of the third SM generation
\begin{eqnarray}
m_{\nu_\tau} << m_{\tau} < m_b << m_t \, ,
\end{eqnarray}
the second assumption leads to the statement that according to the
flavor democracy, the fourth SM family should exist.
In terms of the mass matrix, the above arguments mean
\begin{eqnarray}
M^0 =\frac{ a \, \eta} {\sqrt{2}}\,
\left( \begin{array} {c c c c}
1 & 1 & 1 & 1\\
1 & 1 & 1 & 1\\
1 & 1 & 1 & 1\\
1 & 1 & 1 & 1\\
\end{array} \right)
\end{eqnarray}
which leads to
\begin{eqnarray}
M^m =\frac{ 4a \, \eta}{\sqrt{2}}\,
\left( \begin{array} {c c c c}
0 & 0 & 0 & 0\\
0 & 0 & 0 & 0\\
0 & 0 & 0 & 0\\
0 & 0 & 0 & 1\\
\end{array} \right)
\end{eqnarray}
where $m$ denotes the mass basis.

Now let us state the third assumption:
\begin{itemize}
\item{The coupling $a/\sqrt{2}$ is between $e = g_W \sin \theta_W$ and
$g_Z = g_W / \cos \theta_W$. }
\end{itemize}
Therefore, the fourth family fermions are almost degenerate, in
agreement with the experimental value $\rho = 0.9998 \pm 0.0008$
[1], and their common mass lies between 320 GeV and 730 GeV. The
last value is close to the  upper limit on heavy quark masses which
follows from the partial-wave unitarity at high  energies [10]. It
is interesting to note that with the preferable value of $ a \approx
\sqrt{2}\,g_W$ the flavor democracy predicts the mass of the fourth generation
to be $m_4 \approx 4 a \eta /\sqrt{2}  \approx 8 m_W \approx 640$ GeV.

The masses of the first three families of fermions, as well as
observable inter-family mixings, are generated due to the small
deviations from the full flavor democracy [7, 11, 12]. The
parametrization proposed in [12] gives the values for the
fundamental fermion masses and at the same time predicts the values
of the quark and the lepton CKM matrices. These values are in good
agreement with the experimental data. In principle, flavor democracy
provides the possibility to obtain the small masses for the first
three neutrino species without the see-saw mechanism [13].

The fourth SM family quark pairs will be produced copiously at the
LHC [16, 17] and at the future lepton-hadron colliders [18].
Furthermore, the fourth SM generation can manifest itself via the
pseudo-scalar quarkonium  production at the hadron colliders [19].
The fourth family leptons will clearly manifest themselves at the
future lepton colliders [20, 21]. In addition, the existence of the
extra SM generations leads to an essential increase in the Higgs
boson production  cross section via gluon fusion at the hadron
colliders (see [22-25] and references therein). This indirect
evidence may soon be observed at the Tevatron.

\section{Implications for the Higgs Production}

The cross section for the Higgs boson production via gluon-gluon
fusion at the Tevatron is given by
\begin{eqnarray}
\sigma(p\bar{p}\to HX)=\sigma_0 \tau_H \int_{\tau_H}^1 {dx\over
x}g(x,Q^2)g(\tau_H/x,Q^2)
\end{eqnarray}
where $\tau_H=m_H^2/s$, $g(x,Q^2)$ denotes the gluon distribution function and
\begin{eqnarray}
\sigma_0(gg\to H)= {G_F\,\alpha_s^2(\mu^2)\over 288 \,\sqrt{2} \,
\pi} \, |I|^2
\end{eqnarray}
is the partonic cross section. The amplitude $I$ is the sum of the
quark amplitudes $I_q$ which is a function of $\lambda_q \equiv
(m_q/m_H)^2$, defined as [26]
\begin{eqnarray}
I_q = \frac{3}{2}\, [4\lambda_q +\lambda_q (4\lambda_q-1)f(\lambda_q)]\,\,,\\
f(\lambda_q) = -4 \,(\arcsin ({1\over \sqrt{4\lambda_q}}))^2  \qquad {\rm
  for} \qquad 4\lambda_q > 1 \,\,\,\, \\
f(\lambda_q) = (\ln \frac{1+
  \sqrt{1-4\lambda_q}}{1-\sqrt{1-4\lambda_q}} -i\pi)^2
\qquad {\rm  for} \qquad 4\lambda_q < 1 \,\,.
\end{eqnarray}
The numerical calculations for the Higgs boson production cross
sections in the three SM family case are performed using the HIGLU
software [27] which includes next to leading order (NLO) QCD
corrections [28]. In HIGLU, CTEQ6M [29] distribution is selected for
$g(x,Q^2)$, the natural values are chosen for the factorization
scale $Q^2(=m_H^2)$ of the parton densities and the renormalization
scale $\mu\,(=m_H)$ for the running strong coupling constant
$\alpha_s(\mu)$.

Quarks from the fourth SM generation contribute to the loop mediated
process in the Higgs boson production $gg\to H$ at the hadron
colliders resulting in an enhancement of $\sigma_0$ by a factor of
$\epsilon\cong |I_t + I_{u_4}+ I_{d_4}|^2/|I_t|^2$. Fig. 1 shows
this enhancement factor as a function of the Higgs boson mass in the
four SM families case with $m_4 = $ 200, 320, 640 GeV. For the extra
SM families we find that b-quark loop contribution increases
$\epsilon$ by $9\%-4\%$ depending on the Higgs boson mass in the
range $100-200$ GeV. In the infinitely heavy quark mass limit, the
expected enhancement factors are 9, 25, and 49 for the cases of
four, five, six generations, respectively. Fig. 2 shows the
enhancement factor $\epsilon$ in the four, five and six SM families
cases where quarks from extra generations are assumed to ber
infinitely heavy whereas $m_t=175$ GeV. We also include the QCD
corrections [28] in the decay of the Higgs boson by using the
program HDECAY [30]. Below we deal with the mass region
$115<m_H<200$ GeV, therefore the formulation of $\epsilon$ with
obvious modifications for five and six SM families cases can be a
good approximation. Theoretical uncertainties in the prediction of
the Higgs boson production cross section originate from two sources,
the dependence of the cross sections on parton distributions
(estimated to be around 10\%) and higher order QCD corrections.

\begin{center}
\begin{figure}[H]
\includegraphics [height=9cm,width=9cm]{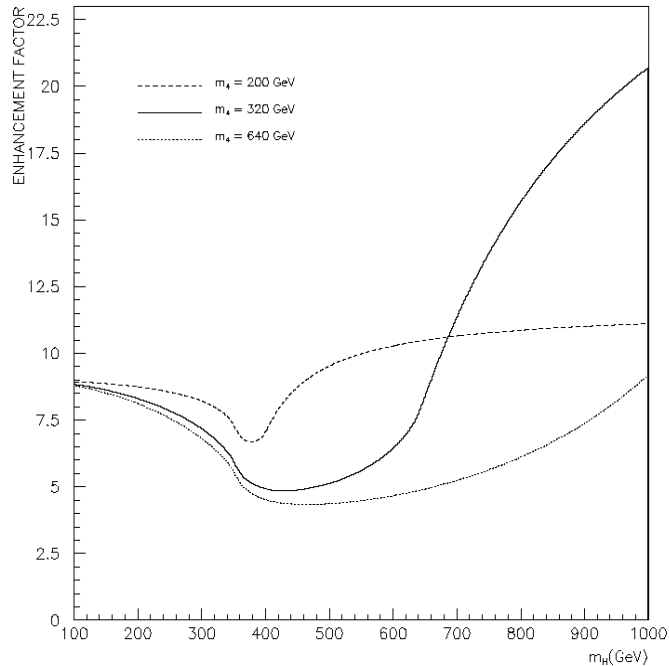}
\caption{The enhancement factor  $\epsilon$ as a function of the Higgs boson mass in
the four SM  families case with $m_4 = $ 200, 320, 640 GeV (upper, mid
and lower curves, respectively).  }
\end{figure}

\begin{figure}[H]
\includegraphics[height=9cm,width=9cm]{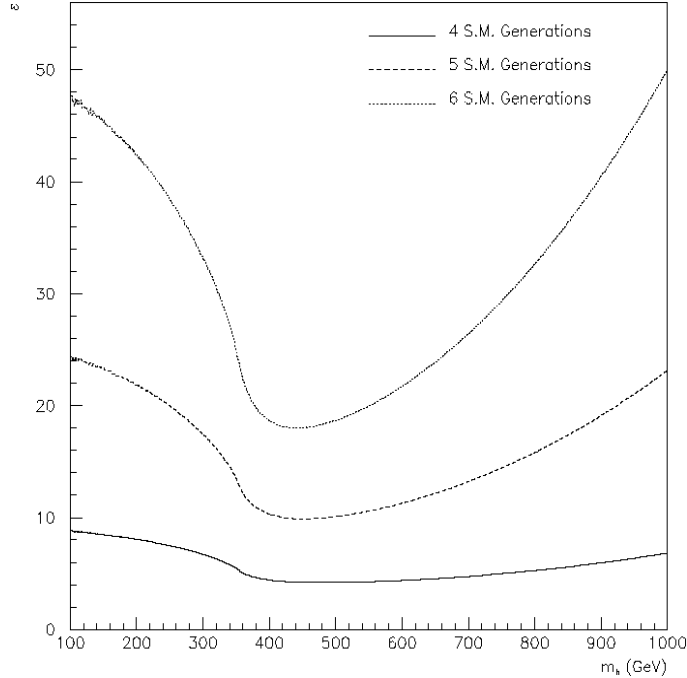}
\caption{The enhancement factor $\epsilon$ as a function of the Higgs boson mass in
the four, five and six SM  families cases where quarks from the extra
generations are assumed to be infinitely heavy.}
\end{figure}
\end{center}

Recently, D0 and CDF collaborations have presented their results on
the search for the  Higgs boson in the channel $H\to WW^{(*)}\to
l\nu l\nu$ [31-36]. Further luminosity upgrade of the Tevatron could
give a chance to observe the Higgs boson at the Tevatron if the
fourth SM family exists.

The Higgs decay width $\Gamma(H\to gg)$ is altered by the presence
of the extra SM generations, due to this effect, the $H \to
WW^{(*)}$ branching ratio changes as shown in Fig. 3. The decay
widths and branching ratios for the Higgs decays are calculated
using HDECAY program [30] after some modifications for extra SM
families. Details on how the branching ratios of all Higgs decay
channels change for extra SM families can be found in [25]. In this
figure 4n, 5n and 6n denote the cases of one, two and three extra SM
generations with neutrinos of mass $\cong 50$ GeV, respectively. We
present the numerical values of the branching ratios depending on
the Higgs boson mass in Table I.  SM-4 and SM-5 denote the extra SM
families with unstable heavy neutrinos, whereas SM-4*, SM-5* and
SM-6* correspond to the extra SM families with $m_{\nu}\cong 50$
GeV. The difference between SM-4 (SM-5) and SM-4* (SM-5*) is due to
additional $H\to\nu_4\bar{\nu}_4$ decay channel in the latter case.

\begin{figure}[H]
\includegraphics[height=9cm,width=12cm]{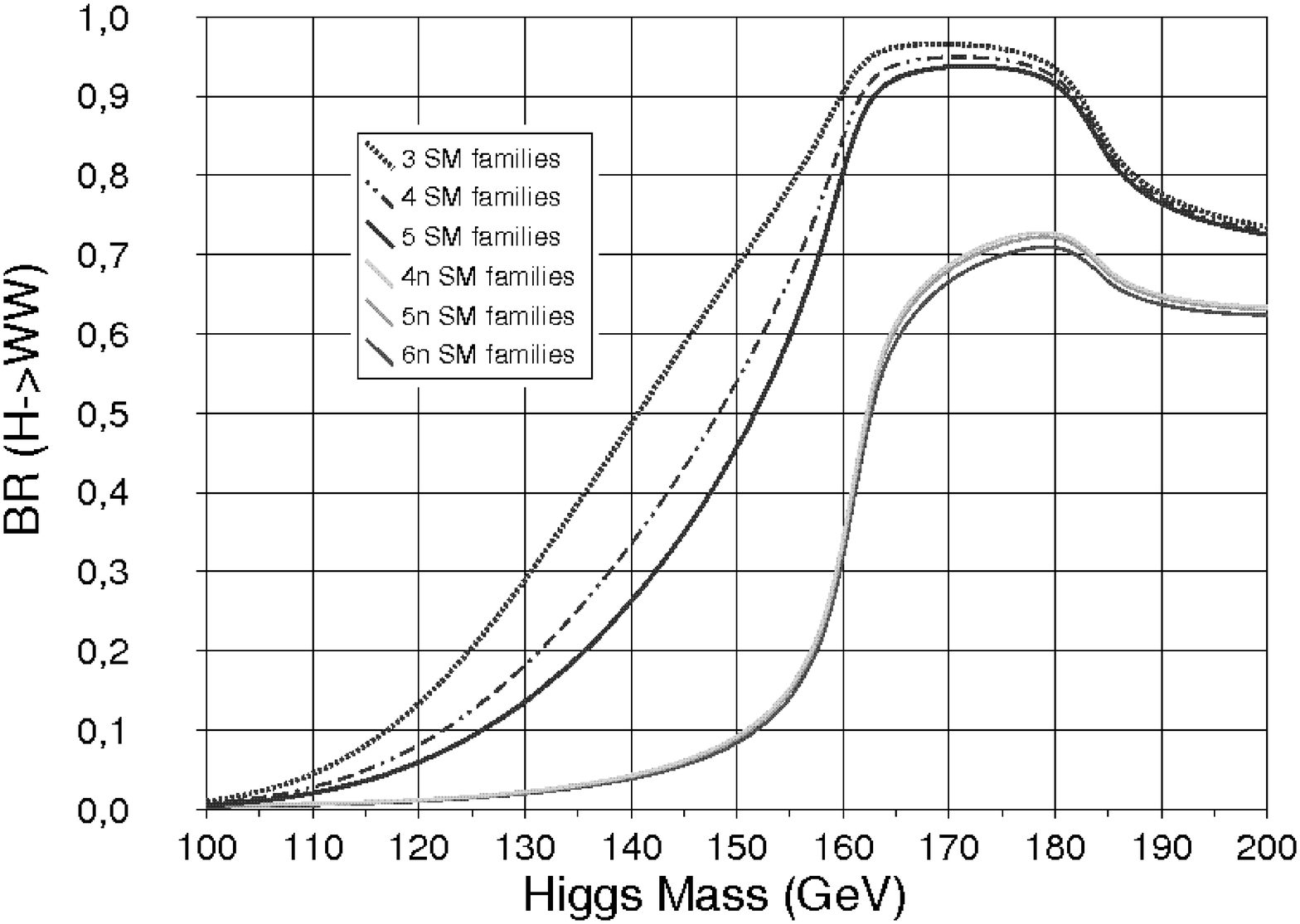}
\caption{The branching ratios for the decay mode $H\to WW^{(*)}$  in various scenarios.}
\end{figure}

\begin{table}[H]
\caption{The branching ratios depending on the mass of the Higgs boson in the
  three, four, five  and six SM families cases. The asterisk denotes
that the calculations are performed assuming $m_{\nu}=50$ GeV for the extra families.}
\begin{tabular}{lllllll}
\hline Mass(GeV)&SM-3&SM-4&SM-5&SM-4*&SM-5*&SM-6*\\
\hline 100&1.02$\times 10^{-2}$&6.73$\times 10^{-3}$ & 5.05$\times
10^{-3}$&6.73$\times 10^{-3}$ &5.05$\times 10^{-3}$ &3.30$\times
10^{-3}$
\\
120 & 1.33$\times 10^{-1}$ & 8.11$\times 10^{-2}$ & 5.95$\times
10^{-2}$ & 1.21$\times 10^{-2}$ & 1.15$\times 10^{-2}$ & 1.03$\times
10^{-2}$
\\
140&4.86$\times 10^{-1}$ &3.35$\times 10^{-1}$ & 2.63$\times
10^{-1}$&4.29$\times 10^{-2}$ &4.15$\times 10^{-2}$ &3.86$\times
10^{-2}$
\\
160&9.05$\times 10^{-1}$ &8.48$\times 10^{-1}$ & 8.05$\times
10^{-1}$&3.43$\times 10^{-1}$ &3.35$\times 10^{-1}$ &3.20$\times
10^{-1}$
\\
180&9.35$\times 10^{-1}$ &9.23$\times 10^{-1}$ & 9.14$\times
10^{-1}$&7.27$\times 10^{-1}$ &7.21$\times 10^{-1}$ &7.09$\times
10^{-1}$
\\
200&7.35$\times 10^{-1}$ &7.29$\times 10^{-1}$ & 7.25$\times
10^{-1}$&6.34$\times 10^{-1}$ &6.31$\times 10^{-1}$ &6.24$\times
10^{-1}$
\\ \hline
\end{tabular}
\end{table}

\section{Results and Conclusions}

In Fig. 4, we added our theoretical predictions for the case of two
extra SM families (SM-5) with unstable heavy neutrinos ($m_{\nu}>
100$ GeV) as well as the possible exclusion limits for the
integrated luminosity $L_{int}=2$ fb$^{-1}$ and 8 fb$^{-1}$.
 It is seen that the  recent Tevatron data excludes SM-5 at
95\% C.L., if the Higgs mass lies in the region 160 GeV $< m_H <$
170 GeV (i.e. this mass region is excluded if there are two extra SM
families with unstable heavy neutrinos). With 2 fb$^{-1}$ integrated
luminosity, the fourth SM family with an unstable neutrino (SM-4)
can be verified or excluded for the region 150 GeV $ < m_H < $ 180
GeV. Similarly, SM-5 can be verified or excluded for the region $m_H
> 130$ GeV with 2 fb$^{-1}$. The upgraded Tevatron is expected to
reach an integrated luminosity of 8 fb$^{-1}$ before the LHC
operation, which means that SM-4 (SM-5) will be verified or excluded
for the Higgs mass region $m_H >$ 140 GeV (120 GeV). However, the
LHC will be able to cover the whole region via the golden mode H
$\rightarrow$ ZZ $ \rightarrow \ell \ell \ell \ell$ and detect the
Higgs signal during the first year of operation if the fourth SM
family exists [25].

\begin{center}
\begin{figure}[t]
\includegraphics[height=10cm,width=10cm]{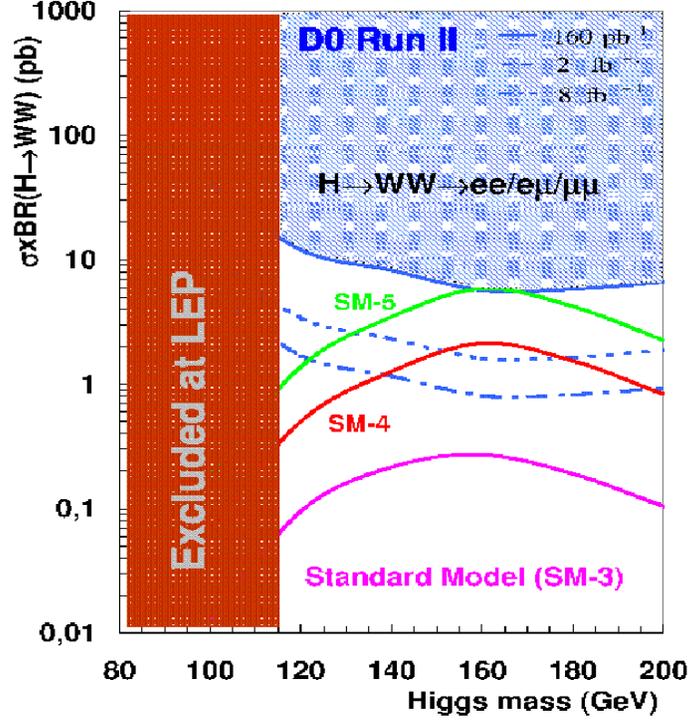}
\caption{The excluded region of $\sigma \times$ BR(H $\rightarrow
WW^{(*)}$) at 95 $\%$ C.L. together with the expectations from the
SM model Higgs boson production and the enhancements due to the extra
SM generations with heavy neutrinos.}
\end{figure}
\end{center}

In Fig. 5, we present our $\sigma\times BR(H\to WW^{(*)})$
predictions for the cases of one, two and three extra SM families
with $m_{\nu}\cong 50$ GeV, SM-4*, SM-5* and SM-6* respectively. If
the Higgs mass lies in the region 165 GeV $< m_H < 185$ GeV,  SM-6*
is excluded at 95 $\%$ C.L.. When $L_{int} = 2 $ fb$^{-1}$ is
reached, the Tevatron data will be able to exclude or verify
SM-6$^*$ (SM-5$^*$) for the mass region $m_H > $ 150 GeV (155 GeV).
With 8 fb$^{-1}$ integrated luminosity, this limit changes to  $m_H
> 145$ GeV (150 GeV) and SM-4$^{*}$ will be observed or excluded in
the range 160 GeV $ < m_H < 195$ GeV.

\begin{center}
\begin{figure}[t]
\includegraphics[height=10cm,width=10cm]{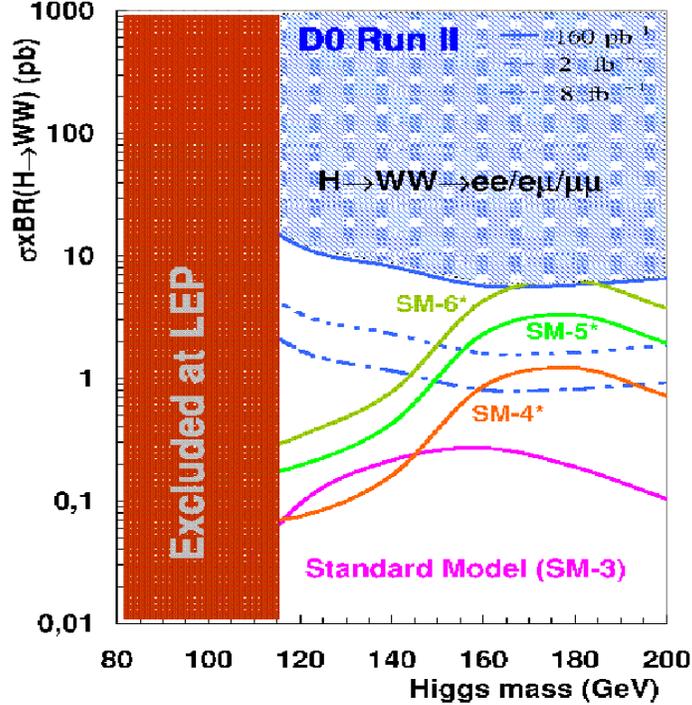}
\caption{The excluded region of $\sigma \times$ BR(H $\rightarrow$ W
W$^{(*)}$) at 95 $\%$ C.L. together with the expectations from the
SM model Higgs boson production and the enhancements due to extra SM
generations with $m_{\nu}=50$ GeV.}
\end{figure}
\end{center}

In Table II, we present the accessible Higgs mass limits at the
Tevatron with $L_{int}=2$ fb$^{-1}$ and 8 fb$^{-1}$ for the extra SM
families.

\begin{center}
\begin{table}[H]
\caption{Accessible mass limits of the Higgs boson at the Tevatron
with $L_{int}=2$ fb$^{-1}$ and 8 fb$^{-1}$ for extra SM families.}
\vskip4mm
\begin{tabular}{lllll}\hline
&&2 fb$^{-1}$&&8 fb$^{-1}$\\
\hline SM-4&&150 $ < m_H < $ 180 GeV&&140 $ < m_H < $ 200 GeV\\
SM-5&&$ > 135$ GeV&&$ >125$ GeV\\
SM-4*&&--&&160 $< m_H<$ 195 GeV \\
 SM-5*&&$>155$ GeV&&$>150$ GeV\\  SM-6*&&$>150$ GeV&&$>145$ GeV
\\\hline
\end{tabular}
\end{table}
\end{center}

Another possibility to observe the fourth SM family quarks at the
Tevatron will be due to their anomalous production via the
quark-gluon fusion process $q g \rightarrow q_4$, if their anomalous
couplings have sufficient strength [37]. Note that the process $q g
\rightarrow q_4$ is analogous to the single excited quark production
[38].

In  conclusion, the existence of the fourth SM family can give the
opportunity to observe the intermediate mass Higgs boson production
at the Tevatron experiments D0 and CDF before the LHC. The fourth SM
family quarks can manifest themselves at the Tevatron as:
Significant enhancement  ($\sim $ 8 times) of the Higgs boson
production cross section via gluon fusion; Pair production of the
fourth family quarks, if $m_{d_4}$ and/or $m_{u_4} < 300$ GeV;
Single resonant production of the fourth family quarks via the
process $q g \rightarrow q_4$.

{\centerline{{\bf Acknowledgments}}}

This work is partially supported by the Turkish State Planning
Organization (DPT) under the grant No 2002K120250, 2003K120190 and
DPT-2006K-120470.

\end{document}